\documentclass{emulateapj}
\shorttitle{WNH stars}
\shortauthors{Smith \& Conti}
\begin{document}

\title{On the role of the WNH phase in the evolution of very massive
stars: Enabling the LBV instability with feedback}

\author{Nathan Smith}
\affil{Astronomy Department, University of California, 601 Campbell
  Hall, Berkeley, CA 94720; nathans@astro.berkeley.edu}

\and

\author{Peter S.\ Conti} 
\affil{JILA, University of Colorado, 389 UCB, Boulder, CO 80309}

\begin{abstract}

We propose the new designation ``WNH'' for luminous Wolf-Rayet (WR)
stars of the nitrogen sequence with hydrogen in their spectra. These
have been commonly referred to as WNL stars (WN7h, for example), but
this new shorthand avoids confusion because there are late-type WN
stars without hydrogen and early-type WN stars with hydrogen.  Clearly
differentiating WNH stars from H-poor/H-free WN stars is critical when
discussing them as potential progenitors of Type Ib/c supernovae and
gamma ray bursts --- the massive WNH stars are {\it not} likely Type
Ib/c supernova progenitors, and are distinct from core-He burning WR
stars in several respects.  We show that the stellar masses of WNH
stars are systematically higher than for {\it bona fide} H-poor WR
stars (both WN and WC), with little overlap.  Also, the hydrogen mass
fractions of the most luminous WNH stars are higher than those of
luminous blue variables (LBVs).  These two trends favor the
interpretation that the most luminous WNH stars are still core-H
burning, preceding the LBV phase (at lower luminosities the WNH stars
are less clearly distinguished from LBVs).  While on the main
sequence, a star's mass is reduced due to winds and its luminosity
slowly rises, so the star increases its Eddington factor, which in
turn strongly increases the mass-loss rate, pushing it even closer to
the Eddington limit.  Accounting for this feedback from mass loss, we
show that observed masses and mass-loss rates of WNH stars are a
natural and expected outcome for very luminous stars approaching the
end of core-H burning (ages $\sim$2 Myr).  Feedback from the strong
WNH wind itself plays a similar role, enabling the eruptive
instability seen subsequently as an LBV.  Altogether, for initial
masses above 40--60 M$_{\odot}$, we find a strong and self-consistent
case that luminous WNH stars are pre-LBVs rather than post-LBVs.  The
steady march toward increased mass-loss rates from feedback also
provides a natural explanation for the continuity in observed spectral
traits from O3~V to O3~If* to WNH noted previously.

\end{abstract}


\section{INTRODUCTION}

The role of mass loss in the evolution of the most massive stars with
initial masses of $\sim$60--150 M$_{\odot}$ is still poorly
understood, partly because the most luminous stars are so rare and
each one seems unique at some level, and partly because accurate
physical parameters are difficult to determine.  Consequently, the
connection between spectral characteristics and evolutionary state is
often mangled.

One persistent mystery concerns the placement of the very luminous
late-type H-rich WN stars (referred to here as ``WNH stars'', as
justified below) in the evolutionary sequence of very massive stars
(see Crowther et al.\ 1995a; Hamann et al.\ 2006; Langer et al.\ 1994).
Because of their high mass-loss rates and consequent emission-line
spectra, they are often discussed alongside or confused with core-He
burning Wolf-Rayet (WR) stars (for reviews of WR stars, see Abbott \&
Conti [1987] and Crowther [2007]), and it is sometimes suggested that
their H content indicates that they represent the early phases of
core-He burning.  In this interpretation, the WNH phase occurs
immediately after -- or sometimes instead of -- the luminous blue
varibale (LBV) phase, marking the beginning of the core-He burning WR
stages (see for example, Schaller et al.\ 1993; Meynet et al.\ 1994;
Maeder \& Meynet 1994).

A different suggestion has been made based on the very high luminosity
of some of the WNH stars and especially based on their membership in
very young massive star clusters like R136 in 30 Doradus, NGC~3603,
and the Carina Nebula (e.g., de Koter et al.\ 1997; Drissen et al.\
1995; Crowther et al.\ 1995a; Moffat \& Seggewiss 1979).  Namely,
these authors suggest that it may be the case instead that the WNH
stars are essentially core-H burning stars that precede the LBV phase.
This suggestion has also been made based on the continuity in spectral
types from O3~V to O3~If* to WNL+h as noted, for example, by
Walborn (1971, 1973, 1974), Walborn et al.\ (2002), Lamers \& Leitherer
(1993), Drissen et al.\ (1995) and Crowther et al.\ (1995a).

This continuity has led to the suggestion that the most luminous stars
may even {\it skip} the LBV phase altogether (Crowther et al.\ 1995a)
-- i.e.\ that they evolve directly from O stars to WR stars (Conti
1976).  However, this direct path is not taken in all cases because it
is contradicted by the existence of very luminous LBVs such as $\eta$
Carinae and the Pistol star.  Langer et al.\ (1994) have discussed the
two possibilities above from the perspective of evolution models, and
they suggest instead that both may be true --- i.e., that a star may
look like a WNH star both before {\it and} after the LBV phase.
Langer et al.\ (1994) investigated a 60 M$_{\odot}$ initial mass
model, and as we will see later, the distinction between stars of 60
and 100 M$_{\odot}$ may be important in terms of their evolutionary
path.

Our main conclusion in this paper is that available evidence strongly
favors the latter interpretation above that WNH stars are pre-LBVs, at
least for the highest luminosities above log(L/L$_{\odot}$)=5.8--6.0.
As discussed below, this is based on the fact that their distributions
of stellar mass are distinct from H-poor stars and that their hydrogen
mass fractions are generally higher than LBVs (although there is a
caveat to this last point depending on initial mass, as discussed
below).  We also argue that one might {\it expect} them to be pre-LBVs
if feedback from mass loss on the main-sequence is accounted for in a
simple way.  In \S 2 we justify the new naming convention ``WNH'', in
\S 3 we examine the masses of WNH stars, and in \S 4 we examine their
hydrogen mass fractions compared to LBVs.  Then in \S 5 we take a look
at how mass loss on the main-sequence might naturally lead to a
WNH-like phase, and how that higher mass-loss phase might drive the
star toward the Eddington limit to become an LBV.  In the last few
sections we discuss further related implications of the scenario where
WNH stars are pre-LBVs.

\section{THE SHORTHAND ``WNH'' TO DESIGNATE A DISTINCT CLASS}

We propose the new designation ``WNH'' for luminous WR-like stars of
the nitrogen sequence exhibiting evidence for hydrogen in their
spectra.  In current practice, these are commonly referred to
collectively as ``WNL stars'', with individual objects classified as
WN6ha, WN7h, etc.\ (see Smith et al.\ 1994; Smith et al.\ 1996;
Crowther et al.\ 1995a).  We do not propose to change this more
specific spectral classification scheme for individual stars.
However, we do advocate that the common usage of ``WNL stars'' as a
group should be changed to ``WNH stars'' for two main reasons, one
being a practical matter and the other having to do with their
evolutionary state and questionable relationship to other WR stars.

The terms WNL, WNE, WCL, and WCE were first used by Vanbeveren \&
Conti (1980) as a shorthand for ``late'' and ``early'' type WR stars
of the nitrogen and carbon sequences.  Its use was quickly adopted as
a convenience by other authors. Later on, the WNL term came to mean
late-type WN stars with hydrogen as most of them have this element
present.  However, observations have firmly established that there are
both late-type WN stars {\it without} hydrogen (the {\it bona fide} WN
stars), and, more problematic, {\it early}-type WN stars with
hydrogen.  A number of early-type (WN3ha, WN4ha, WN5ha) WN stars with
hydrogen have been identified recently in the SMC (Foellmi et al.\
2003a; Foellmi 2004), and some in the LMC as well (Foellmi et al.\
2003b).  One such WNE star with hydrogen is the famous object HD5980,
an eclipsing massive binary where the hydrogen-rich component had been
classified as WN6h, WN11, LBV, or B1.5 Ia at different times during
its LBV-like outburst in the 1990s (Koenigsberger 2004), but had been
classified as WN4h before that.  WN5h is arguably on the border
between early- and late-type, and several of these have been seen in
30 Doradus and Westerlund 1 (Crowther \& Dessart 1998; Crowther et
al.\ 2006).  Other Galactic examples are WR3 (WN3ha), WR10 (WN5ha),
WR48c (WN3h+WC4), WR49 (WN5h), WR109 (WN5h), WR128 (WN4h), and WR152
(WN3h) (Marchenko et al.\ 2004; Hamann et al.\ 2006; van der Hucht
2001).  Clearly, the phenomenon of hydrogen in WN stars is not limited
to late types, rendering ``WNL'' an inappropriate designation as it
refers mainly to WN7/WN8 (Moffat \& Seggewiss 1979).  The shorthand
``WNH'' would encompass both early- and late-type WN stars with
hydrogen.

Second, the designation ``WNH'' is also useful to emphasize the
apparent fact that while they exhibit WR-like spectral features
because of their strong winds (they are sometimes referred to as
``O-stars on steroids''), the WNH stars are distinct from core-He
burning WR stars in several observed characteristics and most probably
in their evolutionary state.  The assertion that they represent a
distinct evolutionary state is justified in the following sections,
where we show that like LBVs, the WNH stars are consistently more
massive than WN and WC stars, and that they have H mass fractions more
in line with LBVs.  We argue that their observed masses, mass-loss
rates, luminosities, wind speeds, and other properties can all be
understood naturally if they represent the later stages of core-H
burning of very massive stars with ages $\sim$2 Myr, making them
immediate precursors of LBVs, or perhaps, quiescent LBVs at lower
luminosities.  Lastly, the high mass-loss rates seen in the WNH phase
might be a {\it necessary condition} to push the star toward the
Eddington limit, again arguing that WNH stars are the best candidates
for the immediate precursors of high luminosity LBVs.

We reiterate that we intend the term ``WNH'' only as collective
shorthand for the class of WN-like stars that show hydrogen in their
spectra, because we believe them to be distinct from H-free WN stars.
While this new terminology is arguably imperfect because one star or
another might challenge categorization or blurr the boundaries, the
term is needed simply to avoid confusion, and it makes a clear
distinction that is also useful for purposes of discussion.  Because
they show a range in H mass fraction (see below), the WNH stars
constitute a somewhat heterogeneous class and may even overlap with
the LBVs.  In that sense, though, the term ``WNH'' has a usefulness
similar to that of ``LBV'', since the LBV class contains stars labeled
variously as S Doradus variables, $\eta$ Car variables, $\alpha$ Cygni
variables, P Cygni stars, Ofpe/WN9 stars, etc., yet we still refer to
them all collectively as ``LBVs''.  The term ``LBV'' has been useful
for broadly discussing stars that we think share a common evolutionary
phase, even though it is not always agreed whether an individual
object is a {\it bona-fide} LBV or not.

The term ``WNH'' is also critical to distinguish these stars from
H-free WN stars, because massive stars are the progenitors of various
types of core-collapse supernovae (SN).  The single most important
observable trait in classifying a SN spectrum is the presence or
absence of hydrogen, making it a Type II or a Type Ib/c, respectively.
The ``H'' in ``WNH'' therefore serves as a clear reminder that
hydrogen is present, and that these are {\it not} to be confused with
the H-free Wolf-Rayet stars that are the likely progenitors of Type Ib
supernovae and possibly also long-duration gamma ray
bursts.\footnote{There are rare objects that confuse the issue, like
SN~2006jc (Foley et al. 2007; Pastorello et al.\ 2007; Smith et al.\
2008), but this only emphasizes the importance of highlighting the
presence of hydrogen.  Also, it is worth noting that some compact
binary scenarios might produce Type Ib/c SNe with lower mass stars
(e.g., Pols \& Devi 2002; De Donder \& Vanbeveren 1998).}  The more
specific classification criteria of WN stars are not relevant or
useful in this context as they mainly describe the temperature and
ionization level of the wind.  Observationally, one cannot determine
if the progenitor of a SN is WNE or WNL, but in pinciple, one can tell
the difference between WN and WNH.

\begin{figure}
\epsscale{1.1}
\plotone{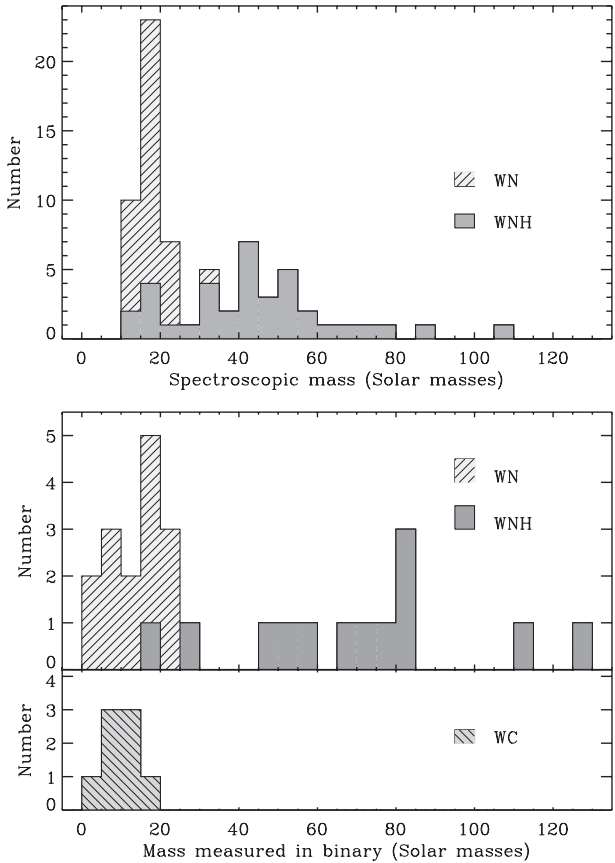}
\caption{Masses of WNH stars compared to H-poor WN and WC stars
  estimated spectroscopically (top) and measured in binary systems
  (bottom).  The spectroscopic masses are taken mostly from Hamann et
  al.\ (2006), except for a few in Carina and 30 Dor from Crowther \&
  Dessart (1998) and de Koter et al.\ (1997).  Masses of WR stars in
  binary systems are taken from van der Hucht (2001), but are
  augmented or superseded in several cases as follows: WR137 (Lefevre
  et al.\ 2005), V444 Cyg (Flores et al.\ 2001), WR141 (Grandchamps \&
  Moffat 1991), WR151 (Villar-Sbaffi et al.\ 2006),
  HD5980A and B (Koenigsberger 2004; Foellmi et al.\ 2008), WR22 (Rauw
  et al.\ 1996), WR20a (Rauw et al.\ 2005; Bonanos et al.\ 2004), WR25
  (Gamen et al.\ 2006); R145/30 Dor and NGC3603/A1 (Moffat 2006,
  priv. comm.).  (Note that in these histograms, numbers of different
  types of objects are ``stacked''.  For example, measured
  in binary systems, there are only 4 WN stars in the 15--20
  M$_{\odot}$ bin, not 5.)}
\end{figure}

\section{MEASURED MASSES OF WNH STARS}

Figure 1 collects mass estimates for WR stars from the literature,
including masses measured both spectroscopically and in binary
systems. It compares present masses of WNH stars to H-poor WN and WC
stars (spectroscopic masses are unreliable for WC stars).  The
spectroscopic masses are taken mostly from Hamann et al.\ (2006), who
derived the masses using the mass-luminosity relation for helium stars
from Langer (1989).  They noted that this relation might not be
adequate for WNH stars, but they also note that when masses measured
in binary systems are available for the same stars for which they
estimated a spectroscopic mass, the two methods show no wild
disagreement.  For example, the spectroscopic mass they derive for the
WNH star WR22 in the Carina Nebula is 74 M$_{\odot}$, compared to 72
M$_{\odot}$ determined from the binary orbit by Rauw et al.\
(1996).\footnote{Although see also Schweikhardt et al.\ (1999) who
derive a mass of 55 $M_{\odot}$ for the WNH component.}  Thus, while
the exact values for the masses in the top panel of Figure 1 might be
somewhat suspect, the main trend of the separation between WN and WNH
stars is not.  Masses measured for WR stars in binary systems are more
reliable, but there are not many of them available, shown in the
bottom panel of Figure 1.  The results for the distribution of masses
for different WR types measured spectroscopically and in binaries show
very good general agreement.

The clear lesson to be learned from Figure 1 is that WNH stars have
systematically higher masses than H-poor WN and WC stars.  With
binaries and spectroscopic masses, the mass distribution of H-poor WN
stars clearly peaks at 15--20 M$_{\odot}$, while WNH stars are spread
more evenly over a large range of masses mostly above 30 M$_{\odot}$.
There are only a few cases of mass overlap between the WN stars and a
small fraction of the WNH stars.  This may be because of WN stars in
binary systems where H is present, or it may signify a real
evolutionary overlap.  Note that the overlap is more prevalent in the
spectroscopic masses where unrecognized binaries are more likely to
contaminate the sample, and where the general shape of the WNH mass
distribution in the 10--25 M$_{\odot}$ range matches that for WN
stars, as one might expect if they are not true WNH stars.  All the
H-poor WR stars have relatively low masses below 30
M$_{\odot}$.\footnote{Under advisement from A.F.J. Moffat (private
comm.) we have excluded WR47, with a mass of 48 M$_{\odot}$ estimated
by Moffat et al.\ (1990), because this system has not been studied in
enough detail yet to determine if H is present or not.  (See also
Karteshiva 2002, who estimate its mass as only 8--12 M$_{\odot}$.)}
The WNH star masses peak around 50 M$_{\odot}$ (for the more numerous
spectroscopic masses), and masses measured in binaries extend to very
large masses -- as high as more than 120 M$_{\odot}$ for the WNH star
R145 in 30 Doradus (Moffat 2006; private comm.).


Overall, then, Figure 1 makes a strong case that WNH stars are in fact
a separate and distinct group of stars from the H-poor WR stars.  It
is obviously logical to assume that WNH stars are more massive than WN
and WC stars because they have not yet shed their H envelopes, and
that they may eventually become H-poor WR stars.  In that case WNH
stars must be at a significantly earlier evolutionary stage, which is
consistent with the fact that they are preferentially seen in massive
clusters within giant H~{\sc ii} regions like the Carina Nebula,
NGC~3603, and 30 Doradus, whereas H-free WN stars are not.  On the
other hand, WNH stars may be consistently more massive than WN or WC
stars simply because they evolve from stars with higher initial masses
and follow different evolutionary paths.

Furthermore, the wide mass discrepancy between WNH and H-poor WR stars
and the fact that the two populations hardly overlap at all provides a
vital clue to their evolutionary states.  It argues strongly {\it
against} the notion that WNH stars are simply the initial stages of
the core-He burning WR phase, where they are still in the process of
shedding the last remaining layers of their hydrogen envelopes,
gradually transitioning into H-free WN stars.  (Besides, shedding the
required 20--80 $M_{\odot}$ in $\la$10$^5$ yr in this transition
requires mass-loss rates higher than observed for WNH stars.)
Instead, the discontinuity in mass of 10's of Solar masses between WNH
and WN stars suggests that some other intermediate stage must quickly
remove the large mass remaining in the hydrogen envelopes of the WNH
stars before they can become WN stars.  The obvious choice for the
subsequent removal of that mass is the violent eruptions of LBVs
(Smith \& Owocki 2006). In other words, based on their masses, WNH
stars are likely to be pre-LBVs, not post-LBV stars.  Interestingly,
unlike the H-poor WR stars, LBVs exhibit a range of masses (with
current masses of roughly 30--100 M$_{\odot}$) that does overlap with
that of the WNH stars, arguing that the two are closely related.

We also note a systematic difference in the masses of WC and WN stars
in Figure 1.  The WC star mass distribution peaks in the 5--15
M$_{\odot}$ range, whereas the WN distribution peaks at higher masses
in the 15--20 M$_{\odot}$ range.  One might expect this if WC stars
are more evolved descendants of WN stars (Paczynski 1973), as the
products of further mass stripping by the powerful WN stellar wind
(Conti 1976) or sudden events like that inferred for the progenitor of
SN~2006jc (Smith et al.\ 2008; Pastorello et al.\ 2007; Foley et al.\
2007).  A prediction of this trend is that SNe of Type Ib should
result from more massive progenitors than Type Ic SNe.  If not, then
an interesting mystery needs to be solved. In any case, the disparity
in mass between the WNH stars and all the other WR types (Fig.\ 1)
supports the main thrust of this paper, and is obviously relevant for
interpreting the progenitors of Types Ib and Ic SNe.

\begin{figure*}
\epsscale{0.93}
\plotone{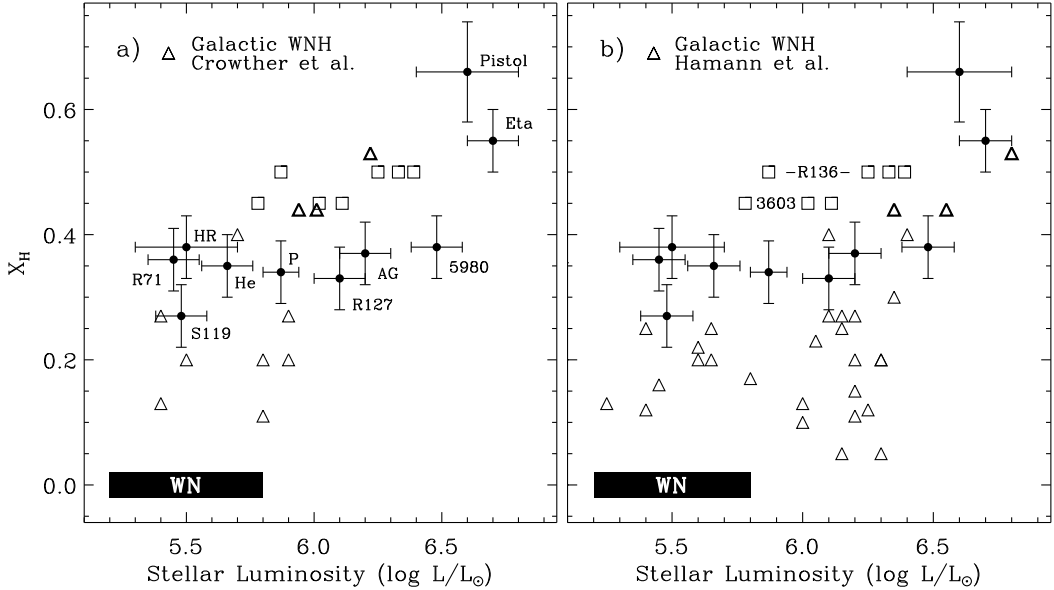}
\caption{ Hydrogen mass fractions, $X_H$, as a function of stellar
  luminosity for WNH stars (unfilled triangles and unfilled squares)
  compared to those of LBVs (filled circles).  The luminosity range of
  H-free WN stars is also shown.  The two panels show $X_H$ and $L$
  for Galactic WNH stars from two different studies: (a) from Crowther
  et al.\ (1995a; or from Crowther, priv.\ comm.), and (b) from Hamann
  et al.\ (2006).  In both studies, the typical uncertainty in $X_H$
  is $\pm$0.05.  For WNH stars common to both studies, the values of
  $X_H$ typically agree to within a few per cent, but the stellar
  luminosities are systematically higher in the study of Hamann et
  al.\ (2006) by roughly 0.5 dex.  This creates a different impression
  for the relationship between WNH stars and LBVs (see text), which is
  why we display the results from both studies.  The luminosities for
  LBVs are the same as in Smith et al.\ (2004), while the values for
  $X_H$ are taken as follows: $\eta$ Car (Hillier et al.\ 2001), the
  Pistol star (Figer et al.\ 1998), HD5980 (Koenigsberger 2004),
  AG~Car, R127, and S119 (Lamers et al.\ 2001), Hen 3-519 and P~Cygni
  (Smith et al.\ 1994), HR~Car (Machado et al.\ 2002), and R71
  (Crowther et al.\ 1995b).  The unfilled squares show the four WNH
  stars in R136 with the luminosities from de Koter et al.\ (1997) and
  a rough value of $X_H\simeq$0.5, since those authors excluded values
  lower than this amount, as well as the three WNH stars in NGC~3603
  with a representative value of $X_H\simeq$0.45, since Drissen et
  al.\ (1995) estimate $X_H$ as 0.4--0.5.}
\end{figure*}

\section{HYDROGEN MASS FRACTIONS}

Figure 2 shows the hydrogen mass fraction, $X_H$, as a function of
stellar luminosity for WNH stars compared to LBVs.  This is obviously
an important quantity for unraveling the evolutionary relationship
between these two classes of stars.  A similar plot and its
implications have already been discussed by Langer et al.\ (1994), but
we update it here with several additional values from the literature.
This updated information is most relevant at the highest luminosities,
as there were only a few data points above log~(L/L$_{\odot}$)=6.0 in
the similar plot by Langer et al.  The behavior of those
high-luminosity objects is the focus here.

Excluding $\eta$ Car and the Pistol star, the LBVs seem to cluster
around $X_H\simeq$0.35, while the WNH stars occupy a large range of
$X_H$ values from 0.05 to 0.5, above and below the values for LBVs.
However, the way that this range of $X_H$ values for WNH stars
compares to those of the LBVs changes with luminosity, and it depends
on which set of WNH luminosities is adopted.  In cases when both
Crowther et al.\ (1995a) and Hamann et al.\ (2006) analyzed the same
target stars, values for $X_H$ generally agree to within a few percent
if they differ at all.  The disagreement is in the stellar
luminosities, with the luminosities from Hamann et al.\ (2006) being
systematically higher by roughly 0.5 dex for the same target stars.
This may be due to the fact that Hamann et al.\ {\it assumed} that WNH
stars had relatively high luminosities.  Shifting the WNH stars
horizontally in Figure 2 significantly impacts our interpretation of
the relative evolutionary status of WNH stars compared to LBVs.
Figures 2$a$ and 2$b$ suggest the following different implications,
respectively:

1) Figure 2$a$ presents a suggestive picture that the relationship
   between WNH stars and LBVs is dependent on luminosity (and hence,
   on initial mass).  Below log~(L/L$_{\odot}$)=5.8 (the upper limit
   for RSGs and for H-free WR stars), we see that the WNH stars are
   generally more evolved with lower hydrogen content than LBVs.  At
   an intermediate luminosity range of log~(L/L$_{\odot}$)=5.8--6.0,
   the case is less clear; WNH stars bracket the $X_H$ values for
   LBVs, suggesting that they could be both pre- and post-LBVs, as in
   the scenario suggested by Langer et al.\ (1994) for a star of
   initial mass 60 M$_{\odot}$ corresponding to P Cygni.  At high
   luminosities above log~(L/L$_{\odot}$)=6.0, however, no WNH stars
   are seen to have smaller hydrogen mass fractions than LBVs -- they
   generally lie above the LBVs, indicating that they should be
   pre-LBVs.

2) With the higher WNH luminosites in Figure 2$b$ (Hamann et al.\
   2006), on the other hand, one would conclude that the LBVs are
   intermixed with and are essentially indistinguishable from WNH stars
   in their abundances, except that LBVs occupy a narrower range.  This
   would imply that WNH stars can be both pre- and post-LBVs, regardless
   of luminosity.  Given the comments above, this is a fair possibility
   at lower luminosities, but seems less likely for the high-luminosity
   stars above log ($L$/$L_{\odot}$)$\ga$5.8.

At this time, we cannot be certain which of these luminosity estimates
to trust, and this highlights the need for accurate estimates of $L$
and $X_H$ in order to understand the evolution of massive stars.
There are a few reasons to favor the somewhat lower luminosities of
Crowther et al.\ (1995a) and the corresponding interpretation implied
by Figure 2$a$.  First, for the Carina Nebula WNH stars WR22, 24, and
25 (shown with bold triangles in Figs.\ 2$a$ and 2$b$), the lower
luminosities of Crowther are in much closer agreement with the
spectroscopically similar WNH stars in R136 and NGC~3603 (de Koter et
al.\ 1997; Drissen et al.\ 1995), shown with squares in Figure 2.
Also, the higher luminosity for WR25 from Hamann et al.\ (2006) makes
this star more luminous than $\eta$~Car, making it difficult to
understand why $\eta$~Car is so much more unstable and more evolved
with a larger N abundance even though they are members of the same
star cluster.  Second, the implication in Figure 2$a$ that WNH stars
precede the LBV phase at high luminosities (above
log~L/L$_{\odot}$=5.8--6.0) is in much better agreement with the
conclusions drawn from considering their current (high) masses
discussed in the previous section, as well as the discussion of the
mass-loss rates to follow.  In Figure 2$b$, on the other hand, the
occurrence of many high luminosity WNH stars with $X_H$ values
significantly lower than LBVs is at odds with the impression that they
typically have higher stellar masses than LBVs.  This argument is
admittedly somewhat circular.  Therefore, we cannot consider the
measured abundances as giving any {\it definite} answer to the
relative evolutionary states of WNH stars and LBVs --- we can only say
that the scenario in Figure 2$a$ fits together in a more consistent
way with the idea that high-luminosity WNH stars are pre-LBVs, while
the implication from Figure 2$b$ that WNH stars are both pre- and
post-LBVs would present an unresolved puzzle (which may nevertheless
be true).  One last reason to favor Figure 2$a$ concerns the
environments in which these stars are found.  The most luminous WNH
stars with higher H content than LBVs reside in massive giant H~{\sc
ii} regions like 30 Dor, Carina, and NGC~3603.  The WNH stars with
lower H conent than LBVs (and at lower $L$) do not reside in such
massive young regions (like most of the LBVs, incidentally).  This
argues for a lower initial mass and larger age for the lower-$L$ WNH
stars.  In any case, further work on the H abundances of WNH stars is
needed.

Of course, factors other than initial mass may influence the apparent
values of $X_H$ as well, such as the star's rotation rate and
consequent level of mixing during core-H burning phases (Maeder \&
Meynet 2000).  This is beyond the scope of our consideration here, but
it obviously may be important.

\begin{figure}
\epsscale{1.13}
\plotone{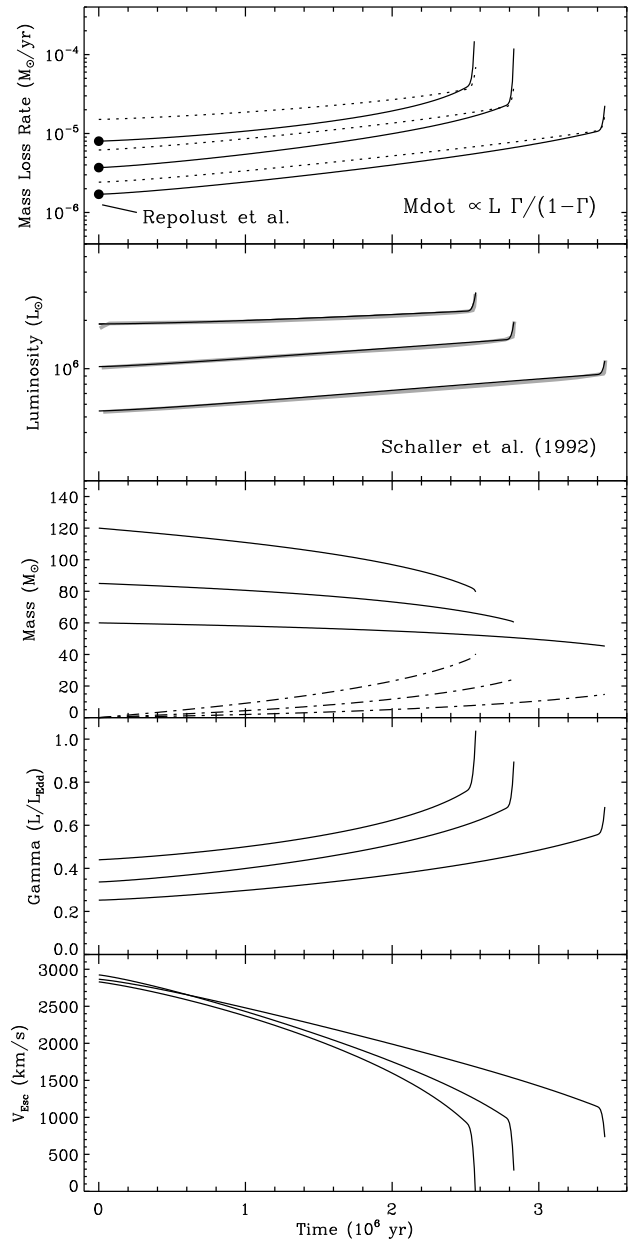}
\caption{Hydrogen core-burning evolution of massive stars with initial
  masses of 120, 85, and 60 M$_{\odot}$, where feedback is included
  such that the mass-loss rate is proportional to
  $L\times\Gamma/(1-\Gamma)$.  Adopted quantities are the luminosity
  during core-H burning from Schaller et al.\ (1992) for each initial
  mass, the initial radius for the corresponding temperature of a
  main-sequence star (only relevant for V$_{\rm esc}$), and the
  initial mass-loss rate.  The initial mass-loss rates (solid dots)
  are those appropriate for the corresponding initial luminosity and
  mass, with moderate clumping factors of $\sim$5 ($\dot{M}$ reduced
  by $\sqrt{5}$), taken from Repolust et al.\ (2004).  For comparison,
  the dotted line in the top panel shows the predicted mass-loss rates
  from Vink et al.\ (2001) for the same $M$ and $L$.  From the
  prescribed initial values, the subsequent mass-loss rate, stellar
  mass, Eddington factor ($\Gamma$=L/L$_{\rm Edd}$), and escape
  velocity are calculated iteratively until the end of core-H burning
  as described in the text.  The plot of stellar mass (solid lines)
  also shows the cumulative mass lost (dot-dashed lines).}
\end{figure}

\section{MASS-LOSS FEEDBACK AND THE EVOLUTIONARY STATES, MASSES, AND AGES OF WNH STARS}

Our primary goal in this paper is to determine how the WNH stars fit
into the evolutionary sequence of massive stars.  When does the WNH
phase ``turn on'', how long does it last, how much mass is lost from
the star, and what are the preceding and subsequent evolutionary
phases?  In the previous sections, we showed that WNH stars are
considerably more massive than H-poor WR stars, arguing that they are
more massive because they have not yet shed their H envelopes and that
they therefore represent an earlier evolutionary phase, before the
heavy mass loss encountered as an LBV.  We also showed that the most
luminous WNH stars tend to be more H-rich than LBVs, arguing again
that they are pre-LBVs.

Another way to attack the problem is to ask when in the lifetime of a
massive star we should {\it expect} the WNH phase to occur, given some
initial mass, luminosity, and mass-loss rate.  These expectations can
then be compared with the measured masses, mass-loss rates,
luminosities, and other properties of WNH stars.


To this end, in the following discussion we consider what the {\it
expected} properties of luminous O-type stars should be as they reach
the end of core H burning.  We are primarily interested in {\it the
rate at which the mass loss rate grows from an initial value during
core-H burning.}  We consider the generic effect of mass loss on the
stellar properties, and on the evolution of the mass-loss rate itself
through a ``feedback'' effect.  The parameterization of mass loss
discussed below is quite simple, and is not claimed to be an adequate
substitute for renewed calculations of stellar evolution codes.
Rather, it is meant only to illustrate the principle that the effect
of mass loss can account for the properties of WNH stars if they are
luminous and massive stars near the end of core H burning.  Our
results argue that renewed efforts to calculate the evolution of
massive stars with lower mass-loss rates are essential in light
of recent observational estimates of lower mass-loss rates due to wind
clumping.  Our arguments here strengthen the case that mass-loss rates
of O-type stars need to revised downward from the ``standard''
observed rates (de Jager et al.\ 1988; Nieuwenhuijzen \& de Jager
1990) due to the observational effects of clumping, and more in line
with (but probably even lower than) the theoretically expected values
of Vink et al.\ (2001).

In Figure 3 we consider the expected properties of stars with initial
masses of 120, 85, and 60 M$_{\odot}$ as a function of time on the
core-H burning main-sequence.  The prescribed inputs here are the
initial mass, the initial value for the mass-loss rate, and the
bolometric luminosity $L(t)$ throughout the core-H burning lifetime.
We adopt $L(t)$ from the Solar metallicity models of Schaller et al.\
(1992), which include mass loss.  Although it would be better to
calculate new stellar evolution models self-consistently instead of
adopting a luminosity from an existing model with different mass-loss
rates, our approach here is a demonstrative first step.  In any case,
the adopted luminosities are sufficient to illustrate the main point
of this paper, which is that the expected climb of the mass loss with
time can account for the apparent properties of the WNH stars.

With these prescribed conditions, Figure 3 shows the time evolution of
the mass-loss rate, the stellar mass, the Eddington factor $\Gamma =
L/L_{\rm Edd}$, where $L_{\rm Edd}$ is the classical Eddington limit
due to electron scattering opacity, and the star's surface escape
velocity.  At each time step, each quantity is calculated iteratively
from the previous time step.  For instance, following the initial
mass, the mass at each subsequent time step is calculated by simply
reducing the mass by $\dot{M} \times \Delta t$.  For reasons that will
become obvious later, we consider only the core-H burning main
sequence lifetime and not core-He burning phases.

The purpose of Figure 3 is to illustrate how these quantities change
during the core-H burning lifetime of the star, in response to the
choice of an initial value for the mass-loss rate.  We are especially
interested in the way that the mass-loss rate grows due to previous
mass loss -- what we refer to below as mass loss ``feedback''.

Essentially all the observable properties of subsequent post-MS phases
and the type of supernova eventually seen depend critically on the
adopted $\dot{M}(t)$.  During core-H burning, stellar evolution
calculations have typically assumed mass-loss rates adopted from
observed ``standard'' values such as those given by de Jager et al.\
(1988) or Nieuwenhuijzen \& de Jager (1990), as was done in Schaller
et al.\ (1992) and subsequent studies, Heger et al.\ (2003), Eldridge
\& Tout (2004), and several others.\footnote{Newer calculations
sometimes use the predicted rates of Vink et al.\ (2001) instead of or
in combination with the de Jager et al.\ (1988) rates, such as Meynet
\& Maeder (2005) and Eldridge \& Vink (2004).}

However, adopting these mass-loss rates is arguably not the best
treatment of mass loss if one is interested in asking what mass-loss
rates to expect as the star evolves, since this method simply
prescribes them (not to mention the fact that $\dot{M}$ values need to
be revised downward due to clumping; see below).  Stars at the same
position in the HR Diagram can have different masses, mass-loss rates,
and other properties, arguing for a different approach.  This may be
one of the reasons that the WNH phase is often assumed to be
associated with later evolutionary phases; for example, following the
end of core-H burning, Schaller et al.\ (1992) {\it impose} a WNH
phase with a constant $\dot{M}$=4$\times$10$^{-5}$ $M_{\odot}$
yr$^{-1}$ until the H envelope is removed.

Here we take a different approach.  Instead of prescribing mass-loss
rates throughout the star's evolution, we consider the effect that
the mass-loss rate for a line-driven stellar wind from a hot massive
star should change during its lifetime, responding to changes in the
star's luminosity with time, as well as to changes in its mass.
During core-H burning, a star's luminosity gradually climbs as the
core contracts (Fig.\ 3), providing one mechanism that will act to
increase the mass-loss rate since the wind is radiatively driven.
Simultaneously, the star's wind is removing mass from the star's
surface, lowering the star's mass considerably, which also acts to
increase the mass loss since the star has a shallower gravitational
potential well.  In essence, both these effects conspire to raise the
star's proximity to the classical Eddington limit, since $\Gamma =
L/L_{\rm Edd}$ is proportional to L/M.  This increase in mass-loss
rate accelerates the growth of $\Gamma$, making the problem worse.
Essentially, this behavior introduces a very important {\it feedback
loop} that is currently not included in stellar evolution
calculations.\footnote{This usage of the term ``feedback'' is
different from that referring to the energy input and metal enrichment
of the ISM by massive stars.}

This feedback effect can be treated in a simple way, sufficient for
our limited purposes here.  The CAK theory of radiatively driven winds
(Castor et al.\ 1975) provides a prescription for how the mass-loss
rate should vary with the star's luminosity $L$ and the Eddington
factor $\Gamma$ (and hence, the star's mass).  Following Owocki
(2003), for example, the dependence can be written as
\begin{equation}
\dot{M} \ \propto \ L \big{(} \frac{\Gamma}{1-\Gamma} \big{)}^{-1 + \frac{1}{\alpha}} 
\end{equation}
where $\alpha$ is the usual CAK power index.  For illustrative
purposes in Figure 2, we adopt $\alpha$=0.5 as is common in
line-driven winds of O-type stars.  Equation (1) then simplifies to
\begin{equation}
\dot{M} \ \propto \ L \big{(} \frac{\Gamma}{1-\Gamma} \big{)}
\end{equation}
which is the mass-loss rate dependence that we adopt in Figure 3.
When this feedback is included, we see that a star's mass-loss rate
will climb steadily and substantially, even during the core-H burning
phase alone.  Given an initial mass-loss rate at $t$=0, one can then
calculate the mass-loss rate and the stellar mass at subsequent times,
self consistently, given $L(t)$, as long as it is safe to assume that
the wind is line-driven.  This assumption will break down as the
Eddington factor climbs near $\Gamma$=1 and the mass-loss rate
skyrockets, when a continuum-driven wind may take over (Smith \&
Owocki 2006; Owocki et al.\ 2004).  Given the simplistic treatment and
the fact that we extrapolate from a single initial value, it is
reassuring that our values of $\dot{M}(t)$ are not too different from
the predicted values of Vink et al.\ (2001; dotted line in the top
panel of Fig.\ 3), especially at the later WNH stages that are the
focus here.

In Figure 3, the initial mass-loss rates we have adopted are the
moderately-clumped rates given by Repolust et al.\ (2004) appropriate
for O-type dwarfs with the adopted luminosity for each of the three
initial masses considered.  These mass-loss rates and the degree of
clumping in stellar winds is a larger issue than we can address here
(see for example, Puls et al.\ 2006; Fullerton et al.\ 2006; Bouret et
al.\ 2005; Smith \& Owocki 2006; Smith 2007a; Eversberg et al.\ 1998;
Lepine \& Moffat 1999).

As a result of the climbing mass-loss rate, we see that $\Gamma$
climbs significantly as well, ramping up at the end of core-H burning.
The implications of this for triggering the LBV instability are
discussed below in \S 7.  Another result of the ramping-up of
$\dot{M}$ is that more of the mass ultimately shed from the star is
lost later in its life, as is also the case if LBV eruptions are a
dominant mode of mass loss (Smith \& Owocki 2006).  This will
significantly impact other properties such as the mass of the He core
produced, as well as the angular momentum evolution of the star and
its rotational mixing.  Lower mass-loss rates will cause the wind to
shed less angular momentum.  In turn, faster rotation will likely
enhance axisymmetric/bipolar mass loss in later evolutionary phases
(Cranmer \& Owocki 1995; Owocki et al.\ 1996, 1998; Owocki \& Gayley
1997; Maeder \& Meynet 2000).  Such effects won't be discussed in
detail below, but they need to be reinvestigated in future stellar
evolution calculations.

\begin{figure}
\epsscale{1.13}
\plotone{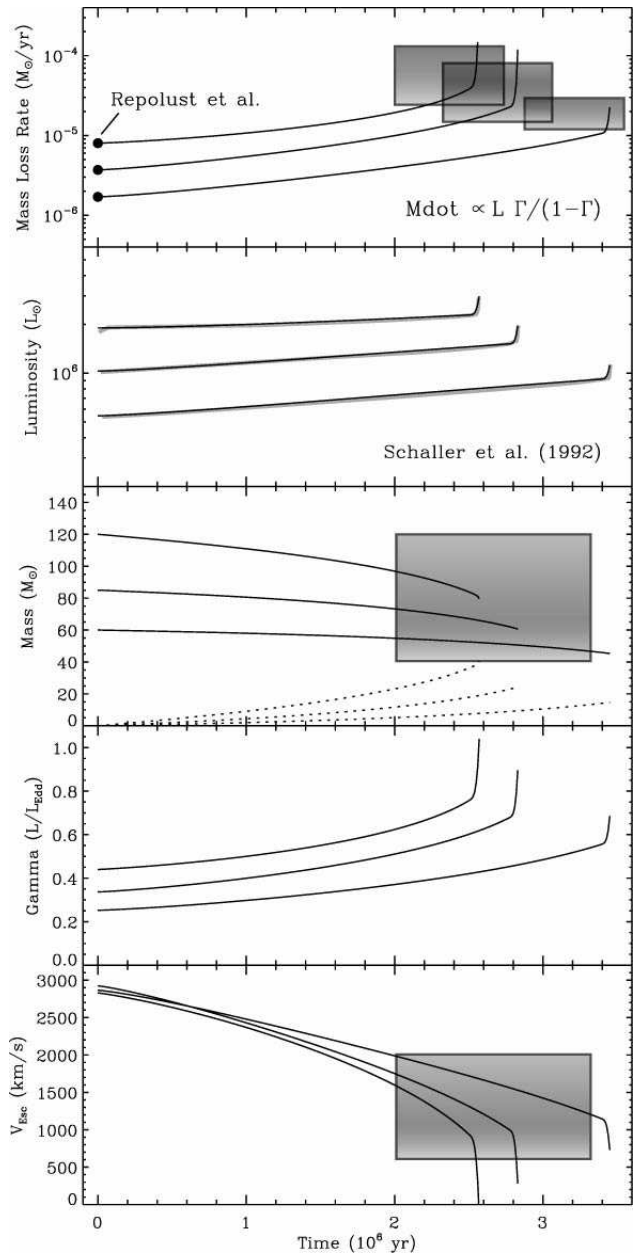}
\caption{Same as Figure 3, but showing some observed values for WNH
  stars for comparison (shaded boxes).  Mass-loss rates and the range
  of terminal wind speeds are taken from Hamann et al.\ (2006).  The
  three boxes in the top panel denote a typical range of mass-loss
  rates corresponding to WNH stars of the corresponding luminosities
  for the three different initial masses treated here.  The WNH
  mass-loss rates from Hamann et al.\ include only weak clumping
  factors (these boxes might be lowered slightly if clumping is more
  severe, but then, so would the corresponding mass-loss rate curves).
  The range of WNH masses from Figure 1 (for the range of luminosities
  considered here) provides less of a constraint, exhibiting a wide
  range of values that are at least consistent with the range of
  predicted masses if WNH stars are near the end of core-H burning.}
\end{figure}

\section{DISCUSSION}

\subsection{Mass-Loss Rates and Ages of WNL Stars}

If Figure 3 gives the expected behavior of $\dot{M}(t)$ for massive
stars, we can then compare it with the observed mass-loss rates of WNH
stars to deduce where they might fit in.  Figure 4 is the same as
Figure 3, but it includes some rough ranges of observed values
corresponding to WNH stars for comparison.

We see here that for these very luminous stars, the mass-loss rate
only climbs to values seen in WNH stars after about 2 Myr from the
beginning of the star's life.  For the luminous WNH stars, typical
mass-loss rates are well above 10$^{-5}$ $M_{\odot}$ yr$^{-1}$
(Crowther et al.\ 1995a).  This favors the interpretation that the WNH
stars have ages of 2--3 Myr.  It is also consistent with the
cohabitation of 3 WNH stars in the same region alongside $\eta$ Car,
whose late evolutionary stage points to an age of roughly 2.5--3 Myr.

An independent check on the ages of WNH stars is given by their
observed wind speeds (bottom panel in Fig.\ 4).  We see that the
observed WNH wind speeds of several hundred to 2000 km/s are not
consistent with ZAMS massive stars, nor with the very fast winds of
H-free WN and WC stars.  Instead, these values are only reached late
in a star's core-H burning lifetime (after roughly 2 Myr) as a natural
consequence of mass loss and the corresponding lowered $g_{\rm eff}$
as the star gets pushed to higher values of $\Gamma$.  Thus, both the
observed mass-loss rates and wind speeds of WNH stars would seem to
favor ages above 2 Myr.

Caveat: One could of course assume that WNH stars are indeed very
young -- that they are somehow born with such high mass-loss rates --
but what should we expect the star's subsequent evolution to look like
in that case?  This is akin to adopting higher mass-loss rates for O
stars (homogeneous winds instead of clumped winds; see Puls et al.\
2006), and this introduces severe problems as discussed next.

\begin{figure}
\epsscale{1.13}
\plotone{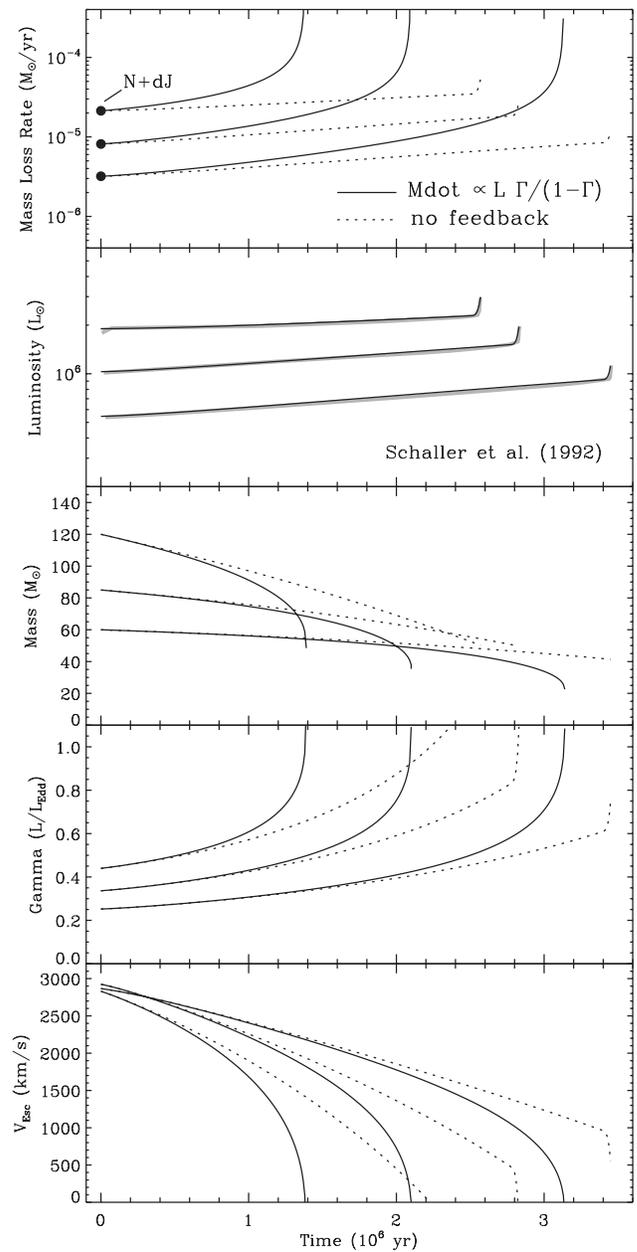}
\caption{Same as Figure 3, but showing how stars run into trouble if
  feedback is included and we adopt the ``standard'' unclumped
  mass-loss rates for the initial state, or if the standard mass-loss
  rates are assumed throughout.  These initial standard mass-loss
  rates are taken from Nieuwenhuijzen \& de Jager (1990), and are only
  a factor of $\sim$2 higher than the moderately-clumped rates of
  Repolust et al.\ (2006) that we used previously in Figures 3 and 4.
  We see that these higher mass-loss rates are clearly not feasible
  because they drive the star to the Eddington limit far too early --
  long before the end of core-H burning for the most massive stars.
  Also shown with dotted lines are the results if we neglect the
  effect of feedback.  This case is arguably artificial, however.}
\end{figure}

\subsection{WNH star masses, and implications for clumped winds and
lowered mass-loss rates of O-type stars}

If WNH stars do indeed reside near the end of core-H burning with ages
$\ga$2 Myr, then their measured masses and luminosities may provide us
with important clues to the mass-loss rates on the main sequence and
the degree of clumping, which is a major issue in stellar evolution.
For example, if WNH stars have relatively high masses when they reach
the ends of their core-H burning lives, then the mass-loss rates
throughout core-H burning can't be very high.  In fact, some very high
masses have been observed for WNH stars.  As shown in Figure 1,
several examples exist of WNH stars with masses of 80--120 M$_{\odot}$
measured in binary systems.

If Figures 3 and 4 are accurate representations of the trend of
$\dot{M}$ on the main sequence, then we can see that the high masses
of WNH stars make sense if they occur at the end of core-H burning ---
{\it but only if the winds are moderately clumped}.  Remember that in
Figure 3 we assumed that the initial mass-loss rates were those of
Repolust et al.\ (2004), which correspond to conservative wind clumping
factors of $\sim$5, reducing the mass-loss rates by factors of $\sim$2
compared to the ``standard'' mass-loss rates\footnote{This is because
of the density-squared nature of the H$\alpha$ and radio continuum
diagnostics of the mass-loss rates.} derived from H$\alpha$ and radio
observations with the assumption of homogeneous winds (de Jager et
al.\ 1988; Nieuwenhuijzen \& de Jager 1990).  In this case, the early
phases with lower mass-loss rates as an O-type star are nearly
irrelevant to the star's total mass loss.  The mass-loss rate
increases later as the star gradually moves into the WNH phase and on
into the LBV phase, indicating that most of a star's mass is lost late
in its lifetime at a quickened pace.  These moderately-clumped
mass-loss rates are just about as high as they can be in order for the
most massive stars to reach the Eddington limit at the end of core-H
burning.

Let's turn this question around and view the problem from another
perspective: If we still include the feedback effect of mass loss
described earlier, as we arguably should, then what will happen if the
stellar winds are {\it not} clumped and we adopt initial mass-loss
rates that are higher?  Figure 5 shows the same type of evolution with
feedback (solid lines) as in Figure 3, with the only difference being
that we started with higher initial mass-loss rates.  These initial
rates (dots) adopt the prescription for the ``standard'' values for
homogeneous winds given by Nieuwenhuijzen \& de Jager (1990).  These
are the values that have most commonly been used in stellar evolution
calculations (e.g., Heger et al.\ 2003), although sometimes rates even
a factor of 2 higher than these are used to match observed statistics
of massive stars (Meynet et al.\ 1994).

The results of adopting the higher mass-loss rates in Figure 5 are
quite dramatic (the same calculations without the feedback effect are
shown with dashed lines in Fig.\ 5, for comparison).  Figure 5 shows
that with these high initial mass-loss rates,\footnote{Note that when
the mass-loss rates climb aggressively, the reduction in the star's
mass is likely to quell the core luminosity somewhat.  Therefore, when
this stage is reached it is likely that our very simple way of
treating the mass loss is invalid, and full stellar evolution
calculations will be needed.  The early push toward those higher
mass-loss rates probably is valid, however.} mass-loss feedback
quickly drives the star's Eddington factor up to $\Gamma$=1 in only
1.3, 2 Myr and 3 Myr for the 120, 85, and 60 M$_{\odot}$ models,
respectively, likely triggering a very early LBV-like phase with
catastrophic mass loss.  However, this is certainly in conflict with
observations, at least for the 85 to 120 M$_{\odot}$ models, since ---
aside from the one case of $\eta$ Carinae --- the stars in massive
young clusters like R136, NGC3603, and Carina typically do not have
very massive shells of recently ejected material around them.  Thus,
we consider it unlikely that very luminous stars are born with
mass-loss rates as high as those seen in WNH stars or with the
mass-loss rates corresponding to homogeneous stellar winds.  Also, the
fact that massive stars reach the end of core-H burning as WNH stars
with relatively high masses cannot be explained by models in Figure 5
with high mass-loss rates because too much mass has already been lost,
whereas the high masses of WNH stars arise naturally for
moderately-clumped winds.  This is the sensitive nature of the
feedback effect -- that even a small reduction by a factor of $\sim$2
in the initial mass-loss rate can have dramatic consequences later in
a star's life.

Furthermore, there's also the case of LBVs like $\eta$ Car to
consider: it is the most luminous and most evolved member of a rich
region containing over 65 O-type stars, as well as the 3 well-known
WNH stars (see Smith 2006 for a review, including details of the age
differences among clusters in the region).  It is fair to assume that
the current LBV phase of $\eta$ Car is not only a post-MS phase, but
probably also a post-WNH phase, since its ejecta are more nitrogen
rich than the WNH stars in Carina.  It is also safe to assume that
$\eta$ Car has advanced further in its evolution sooner than the WNH
stars of the same age in this region simply because it is more
luminous and started with a higher initial mass.  Now, $\eta$ Car is
seen today surviving as a very massive star of around 100 M$_{\odot}$
or more (allowing for a hypothetical $\sim$30 $M_{\odot}$ companion
star), and we measure a total of something like 20-35 M$_{\odot}$ in
its circumstellar material ejected in only the last few thousand years
(the Homunculus, plus more extended outer material; see Smith et al.\
2003, 2005).  That means $\eta$ Car began its LBV phase -- and {\it
ended} its MS and/or WNH phase -- with more than 120 M$_{\odot}$ still
bound to the star.  If there really is an upper limit of about 150
M$_{\odot}$ to the mass of stars (Figer 2005), then this mass budget
demands that the O-star and WNH winds were indeed quite meager before
reaching this phase.  Similarly, there's the Pistol star to consider
as well, which is also a post-MS object and has a present-day mass
that probably exceeds 100 M$_{\odot}$ (Figer et al.\ 1998).

In conclusion, then, the relatively high masses of WNH stars, the high
masses of the most luminous LBVs, and the intuition that we should
include the feedback effect of mass loss all argue that line-driven
winds of massive stars {\it must} be clumped.  This argument is
independent of the many spectroscopic clues that these winds are
clumped (e.g., Puls et al.\ 2006; Bouret et al.\ 2005; Fullerton et
al.\ 2006; Eversberg et al.\ 1998; Lepine \& Moffat 1999). We find
that the mass-loss rate reductions due to clumping must be at least a
factor of 2 compared to ``standard'' rates for homogeneous winds,
consistent with the conservative factors adopted by Repolust et al.\
(2004).  It is encouraging that this amount of mass-loss reduction
brings the mass-loss rates of O stars into better agreement with
theoretical predictions for line-driven winds of O-type stars (e.g.,
Vink et al.\ 2001), although those may still be to high.

\subsection{WNH stars as pre-LBVs, not core-He burning WR stars}

Unfortunately, it is difficult to reliably determine the age of a WNH
star directly from observations of its environment.
One can easily deduce cluster ages from observations that are either
too low or too high by 1 Myr or more: The essential problem is that
the lifetimes of these very massive stars are so short that they are
often comparable to the uncertainty in the age of their parent
cluster.  In addition, that parent cluster or association may have a
real age spead that makes the problem even worse.  Thus, identifying a
WNH star in a young 1--2 Myr cluster like R136 or NGC3603 (if their
ages are that low) does not necessarily mean that particular WNH star
itself has an age of 1--2 Myr.  In any case, the validity of ascribing
a single-valued age to a given cluster is highly debatable.

Since we can't really trust direct estimates of the ages for
individual WNH stars based on their environments -- at least not at
the precision of $\pm$1 Myr -- we must try to infer relative ages in
other ways.  In this paper we have demonstrated that there are four
main reasons to favor the interpretation that luminous WNH stars are
pre-LBVs:

1) The stellar masses of WNH stars are systematically higher than
   H-poor WN stars, and more specifically, they have very different
   distributions (Fig.\ 1).  Namely, H-free WR stars are all heavily
   clustered around 15--20 M$_{\odot}$, whereas WNH stars seem to be
   spread evenly from 20 M$_{\odot}$ all the way up to the most
   massive stars known well above 100 M$_{\odot}$.  This is clearly in
   conflict with the notion that WNH stars are in the process of
   continuously becoming WN stars through {\it their own} stellar
   winds.  Instead, the strong discontinuity in mass distribution
   between the WNH and WN stars argues for an intervening phase of
   episodic mass loss that quickly sheds 10's of Solar masses and removes
   essentially all the remaining H envelope.  This intervening rapid
   mass-loss phase is almost certainly the LBV phase (regardless of
   what the interpretation for the cause of the LBV phase might be;
   i.e.\ inherent instability of single stars vs. binary mergers,
   etc.).  Unlike for H-poor WN stars, the stellar masses for LBVs
   overlap quite well with the WNH stars.

2) At high luminosities, the hydrogen mass fractions, $X_H$, for WNH
   stars tend to be higher than for LBVs (Fig.\ 2$a$).  This requires
   that they are less evolved.  We noted that this case is not
   definitive, since the higher WNH luminosities from Hamann et al.\
   (2006) paint a somewhat different picture than the lower
   luminosities of Crowther et al.\ (1995a).  However, the properties
   of WNH stars in the cores of NGC~3603 and 30 Doradus would seem to
   favor the interpretation that they are pre-LBVs based on their
   abundances (de Koter et al.\ 1997; Drissen et al.\ 1995).

3) We showed that if one includes feedback due to mass loss on the
    main sequence, then starting with a moderately-clumped initial
    mass-loss rate, one naturally {\it expects} the star's mass-loss
    rate to climb after about 2 Myr to values that would make it
    appear as a WNH star, near the end of core-H burning and before
    the LBV phase (Figs.\ 3 \& 4).  As we showed earlier, if massive
    stars are born with the high mass-loss rates of WNH stars, then
    the subsequent evolution does not make sense (Fig.\ 5).

4) Conversely, if the mass-loss rates are lower, then the effect of
    feedback is not so severe. Interestingly, the relative {\it rate}
    at which $\dot{M}$(t) climbs with the lower initial mass-loss
    rates of clumped winds (Fig.\ 3) more closely matches that of the
    {\it observed} mass-loss rates, even though the observed rates are
    offset to higher values.  Therefore, simply lowering the observed
    ``standard'' values of Nieuwenhuijzen \& de jager (1990) by a
    factor of 2--3 gives a fairly accurate match to the expected
    mass-loss rates with feedback and clumping.  This provides a
    powerful, self-consistent argument that the mass-loss rates are in
    fact lowered due to clumping in stellar winds.

One of the interesting results of Figure 3 is that this steady march
toward increased mass-loss rates from feedback on the main sequence
also provides a natural explanation for the apparent continuity in
observed spectral traits from O3~V $\rightarrow$ O3~If* $\rightarrow$
WNH noted previously (Walborn 1971, 1973, 1974; Walborn et al.\ 2002;
Walborn \& Blades 1997; Conti 1976; Melnick 1985; Massey \& Hunter
1998; Lamers \& Leitherer 1993; Drissen et al.\ 1995; Crowther et al.\
1995a; etc.), and onward from WNH $\rightarrow$ LBV as well.  This
sequence is known to show intermediate stages, such as hot slash stars
like Melnick~42 in 30 Dor and weak-lined WNH stars like WR25 (e.g.,
Walborn et al.\ 1992), attributed mainly to changes in wind density
during stellar evolution.  We argue that no special circumstances like
pulsationally-enhanced mass loss, rapid rotation, binary mergers, or
unusual abundances are needed to account for the presence of WNH stars
in massive young clusters with ages of around 2--3 Myr --- it is a
natural outcome of intially moderate mass loss on the main sequence
that gradually grows more severe later on (Fig.\ 3).

\subsection{And Where are the Luminous Post-LBVs?}

There are no H-free WR stars with luminosities above
log~(L/L$_{\odot}$)=5.8 (Fig.\ 2), and if we favor the results shown
in Figure 2$a$, there are not even any WNH stars with lower hydrogen
mass fractions than LBVs for luminosities above
log~(L/L$_{\odot}$)=6.0.  One possibility is that the LBV mass loss is
so extreme that giant eruptions can completely remove the remaining H
envelope to expose the bare He core.  Thus, in Figure 2, a star would
effectively move instantaneously from the position of an LBV like
AG~Car to a lower-luminosity WR star.  However, the clear absence of
H-free WR stars above log~(L/L$_{\odot}$)=5.8 is puzzling in that
case, as is the general dearth of H-free WN stars in giant H~{\sc ii}
regions.  The sudden removal of the outer layers at the end of core-H
burning should not much affect the luminosity of the He core that
remains behind, so where are these luminous H-free stars?  One way out
of this predicament, hinted at by Figure 2$a$, could be if the more
massive stars explode before shedding their H envelopes.

In fact, there is mounting evidence that some massive stars may
explode during the LBV phase before ever making it all the way to the
H-free WR stage (see the discussions in Smith \& Owocki 2006; Smith
2007a, and Gal-Yam et al.\ 2007).  Some examples are the recent Type
IIn event SN~2006gy, which may have been the explosion of a very
massive star like $\eta$ Carinae (Smith et al.\ 2007), the Type IIn
event SN~2006gl, whose putative progenitor star identified by Gal-Yam
et al.\ (2007) had photometric properties consistent with an LBV, the
variable radio properties of some SNe (Kotak \& Vink 2006; for other
interpretations, however, see Soderberg et al.\ [2006] and Ryder et
al.\ [2006]), the SN1987A-like nebula around the LBV star HD168625 (Smith
2007b), plus many other Type IIn SNe with dense environments.  There
are also some He-rich stars (perhaps LBV/WN transition stars) that
appear to have suffered LBV-like mass ejections shortly before a Type
Ib/c SN explosion -- the clearest example being SN~2006jc (see Foley
et al.\ 2007; Pastorello et al.\ 2007; Smith et al.\ 2008), which was
actually {\it observed} to have an LBV-like event 2 yr prior to the SN
explosion.  For the low-luminosity LBVs like HD168625 or R71,
explosion as an LBV is not necessarily a problem -- or even a surprise
-- because those lower-L stars are likely to be in a post-RSG phase
(see Smith et al.\ 2004).  For the high-luminosity LBVs above
log~(L/L$_{\odot}$)=5.8, however, it presents a serious challenge to
our current paradigms of stellar evolution.

\subsection{And What Have We Left Out?}

Obviously, we have stopped short of a full calculation of stellar
evolution models that would take into account the way that the core
luminosity may respond to mass loss.  However, in exploring the
feedback effect, we adopted the core luminosities from the models of
Schaller et al.\ (1992), which used relatively high mass-loss rates as
noted earlier.  Therefore, with the lower initial mass-loss rates we
argue for here (and a consequent higher core luminosity), the feedback
effect we propose could be even more extreme.  Nevertheless, we only
set out to demonstrate the principle of the feedback effect and that
it can lead to high mass-loss rates of WNH stars if they are in the
latter part of core-H burning for initially very massive O-type stars.
Renewed efforts to calculate full stellar evolution models are
encouraged.

Our simple analysis does not account for chemical mixing and possible
effects of rotation on that parameter, which is still a central
problem in the evolution of massive stars.  It is quite possible that
rotationally-enhanced mixing could lead stars with identical masses to
evolve on somewhat different paths, hence surface abundances could
differ. Thus, the exact time of ``onset'' of the WNH phase might vary
from one star to another -- even for stars of the same initial mass --
based on the initial rotation rate and efficiency of the mixing.
Therefore, our quoted age of $\sim$2 Myr for the onset of the WNH
phase, which is imprecise to begin with, is not meant to be
definitive.

It is well known that LBVs can undergo minor outbursts (i.e., ``S Dor
eruptions'') where they may change their mass-loss rate while
remaining at constant bolometric luminosity, causing a corresponding
change in apparent temperature and wind speed due to a pseudo
photosphere.  More relevant here is that they can also suffer giant
eruptions with violent sudden mass loss, like those seen in $\eta$ Car
and P Cygni (see, e.g., Davidson et al.\ 1989), where they may eject
$\sim$10 $M_{\odot}$ in a single burst (Smith et al.\ 2003; Smith \&
Owocki 2006), and these event are likely to repeat.  There is a
suspicion in the hot star community that the most luminous objects
(e.g. $\eta$ Car or the Pistol star) might have fewer but more severe
outbursts, while less luminous LBVs could repeat these episodes many
times with each individual event less violent than for $\eta$ Car
(again see, e.g., contributions in Davidson et al.\ 1989).  The mass
lost each time, the frequency, and the total number of such outbursts
as a function of luminosity or intial mass, combined with the
potential regulating/perturbing effect of close binaries, are
parameters that are badly needed for modeling the late evolution of
massive stars, but we are still far from accurate empirical
prescriptions of this mass loss behavior.  Therefore, we must remain
skeptical of the predictions of stellar evolution codes beyond the end
of core-H burning for very massive stars.

How do the WNH stars fit into this scenario?  Could a WNH become an
LBV and thence return to the WNH stage again?  This is thought to be
the case, specifically, for some Ofpe/WN9 stars, since the LBVs AG Car
and R127 both look like Ofpe/WN9 stars in their quiescent state (Stahl
1986; Stahl et al.\ 1983).  Perhaps something similar is happening for
HD5980 in the SMC.  Can we determine if such LBV/WNH transition stars
are occurring relatively early or late in a broader LBV phase?  How
does the initial stellar composition affect the WNH/LBV scenario we
propose here?  Does the presence of a number of early WN stars in the
SMC that are also WNH tell us something?

\section{SUMMARY: THE STRONG WINDS OF WNH STARS ENABLE THE LBV PHASE}

We have shown that the masses, mass-loss rates, and abundances of
luminous WNH stars are distinct from H-free WR stars, that they can be
explained naturally if they are in the late stages of core-H burning,
and that this can be understood as a direct consequence of mass-loss
feedback during core-H burning.  We treat this feedback very
simplistically as the dependence of the mass-loss rate on the
luminosity and Eddington factor associated with conventional
line-driven wind theory, as explained in \S 5.  We have ignored
additional effects such as rotationally or pulsationally enhanced mass
loss, but those effects may become necessary if O-star mass-loss rates
are indeed reduced much below the values corresponding to moderate
clumping factors (Repolust et al.\ 2004) that we have adopted here.
The feedback causes a steady climb in the star's mass-loss rate.

Just as this O-star mass loss provides a sort of feedback that leads
to the higher mass-loss rates of WNH stars, so too does the higher
mass-loss rate of the WNH phase enable further instability in later
phases.  Namely, with increased mass loss, the WNH wind lowers the
stellar mass even further as the luminosity continues to climb at an
ever faster pace.  This runaway eventually pushes the star to the
classical Eddington limit ($\Gamma$=1) as shown in Figure 3.  At some
point along the way, perhaps at $\Gamma\simeq$0.9, an opacity modified
Eddington limit (e.g., Lamers \& Fitzpatrick 1988; Appenzeller 1986;
and many contributions in Davidson et al.\ 1989) presumably takes over
and runaway mass-loss will ensue, marking the beginning of the LBV
phase.

Exactly when this is initiated relative to the brief transition from
core H to He burning is unclear, and would seem to vary with the
initial mass of the star and with metallicity.  For the most massive
stars, the sharp increase in luminosity right at this transition
pushes the star past $\Gamma$=1.  However, down near 60 M$_{\odot}$,
for instance, it would appear that the star can linger on well into
core He burning before actually triggering the LBV instability.
Again, this is very interesting from the point of view of LBVs being
potential Type IIn SN progenitors.

Another consequence of lower mass-loss rates throughout
core-H burning is that the star will suffer less angular momentum
loss.  This, in turn, makes rotation have an even more important
effect in the later stages as an LBV, as discussed by Langer and
others (Langer 1998; Langer et al.\ 1994; Glatzel 1998; Maeder \&
Meynet 2000).  Perhaps this rotation is critical for triggering the
LBV instability for some range of LBVs, since models for initial
masses below 85 M$_{\odot}$ do not quite reach $\Gamma\ga$1 by the end
of core-H burning (Fig.\ 3).  In any case, what happens to the
structure of the star after the onset of the LBV phase has not yet
been adequately addressed in stellar evolution calculations, because
short duration eruptions and explosions of LBVs seem to be an integral
part of exceeding the Eddington limit.

\acknowledgments \scriptsize

We have benefitted from numerous discussions about stellar winds and
WNH stars with Paul Crowther, Gloria Koenigsberger, Tony Moffat, Stan
Owocki, Joachim Puls, and Jorick Vink. We are particularly grateful to
P.\ Crowther, T.\ Moffat, Nolan Walborn, and an anonymous referee for
detailed comments on the manuscript.  NS acknowledges support from
NASA through STScI and JPL, and PSC appreciates support from the NSF.


\end{document}